\documentclass[12pt,a4]{article}
\usepackage{latexsym,epsfig,amssymb,euscript,amsmath,verbatim,cite,subfigure}
\usepackage{float}
\usepackage{verbatim} 
\newcommand{\be}{\begin{equation}}
\newcommand{\ee}{\end{equation}}
\newcommand{\bea}{\begin{eqnarray}}
\newcommand{\eea}{\end{eqnarray}}

\numberwithin{equation}{section}
\usepackage{bm}
\usepackage{hyperref}

\usepackage[usenames,dvipsnames]{color}

\begin{document}

\pagestyle{empty}
\vspace{1.8cm}

\begin{center}
{\LARGE{\bf  {Bounds on charge and heat diffusivities in momentum dissipating holography}}}
\vspace{1cm}

{\large{Andrea Amoretti$^{a,d,}$\footnote{\tt andrea.amoretti@ge.infn.it },
Alessandro Braggio$^{b,}$\footnote{\tt alessandro.braggio@spin.cnr.it }, \\ 
Nicodemo Magnoli$^{a,}$\footnote{\tt nicodemo.magnoli@ge.infn.it},
Daniele Musso$^{c,}$\footnote{\tt dmusso@ictp.it} 
\\[1cm]}}

{\small{
{}$^a$  Dipartimento di Fisica, Universit\`a di Genova,\\
via Dodecaneso 33, I-16146, Genova, Italy\\and\\I.N.F.N. - Sezione di Genova\\
\medskip
{}$^b$  CNR-SPIN, Via Dodecaneso 33, 16146, Genova, Italy\\
\medskip
{}$^c$ Abdus Salam International Centre for Theoretical Physics (ICTP)\\
Strada Costiera 11, I-34151 Trieste, Italy\\
\medskip
{}$^d$  Lorentz Institute for Theoretical Physics\\ Niels Bohrweg 2, Leiden NL-2333 CA, The Netherlands
}}
\vspace{1cm}

{\bf Abstract}

Inspired by a recently conjectured universal bound for thermo-electric diffusion constants in
quantum critical, strongly coupled systems and relying on holographic analytical computations, 
we investigate the possibility of formulating Planckian bounds in different holographic 
models featuring momentum dissipation.
For a simple massive gravity dilaton model at zero charge density we find robust 
linear in temperature resistivity and entropy density alongside a constant electric 
susceptibility. 
In addition we explicitly find that the sum of the thermo-electric 
diffusion constants is bounded.


\end{center}

\newpage
\setcounter{page}{1} \pagestyle{plain} \renewcommand{\thefootnote}{\arabic{footnote}} \setcounter{footnote}{0}
\section{Introduction}

The description of universal behaviors
of strongly correlated systems despite the differences in their microscopic 
structure is challenging and of high physical relevance.
One of the best known examples is the linear-in-temperature resistivity regime
shared by many apparently very different materials like cuprates, heavy fermions, 
pnictides, ruthenates and fullurenes \cite{bruin}.
In recent times, a great effort has been put into characterizing the universal behavior
of these materials by determining bounds on appropriate physical quantities. 
The reason why this approach is promising is twofold: first, a bound can 
be formulated independently of the microscopic detail of the system and can therefore aspire to be universal;
secondly, in an incoherent quantum critical regime, the system equilibrates on a time scale
independent of any microscopic energy scale and just determined by the temperature. 
Therefore, as a consequence of the insensitivity to energy scales other than the 
temperature, incoherent strongly correlated systems may tend to saturate general temperature-dependent bounds.

Strongly correlated models amenable to quantitative theoretical control 
are extremely rare. In this respect, the gauge/gravity duality is very useful since it 
provides a dual description for some strongly interacting quantum field theories where 
equilibrium and transport properties can be computed. 
Actually, the celebrated proposal of a universal bound for momentum diffusion 
in strongly coupled systems was advanced relying on holography. 
This bound was originally conjectured in \cite{Kovtun:2004de} in 
terms of the ratio of the shear viscosity to the entropy density, namely 
$\eta/s \ge \hbar/(4 \pi k_B)$. 
The approximate saturation of the viscosity bound has been 
subsequently observed in the quark gluon plasma created at the heavy ion colliders, as well
as in the unitary fermionic cold atom gas \cite{Adams:2012th}.
Moreover, recent experimental ARPES measurements on an optimally doped cuprate show that, 
also in this case, the ratio $\eta/s$ due to the intrinsic electronic contribution is close to saturate the afore mentioned bound \cite{rameu}.
This experimental evidence strongly suggests that the transport properties of the strange metals due to
the strongly correlated electron behavior could be described in the holographic framework.

The physical origin of the $\eta/s$ bound can be explained 
referring to the concept of ``Planckian'' dissipation \cite{sachdevb,zaanen1}: the fact that, in the finite 
temperature quantum critical state, physical quantities relax extremely rapidly and the equilibration 
rate is substantially determined by the uncertainty principle, namely $\tau_{\text{eq}} \sim \hbar/(k_B T)$. 
Recent experimental observations showing that a wide range of strange metals in the linear-in-temperature resistivity regime present an equilibration rate of this kind \cite{bruin} 
strongly supports the idea that a unified framework might capture the properties of these strongly 
interacting materials.
 
Relying on a wider applicability of similar arguments, it is tempting to find whether other physical
observables in strange metals and, more generally, in strongly correlated materials have to saturate
a bound due to Planckian dissipation.
However, the identification of the correct kinematic framework where one can formulate universal bounds 
is, in general, not easy. The reason being that in metallic materials the mechanisms for momentum relaxation, 
such as the lattice or scattering from disorder, which are necessary to make the conductivities finite, 
are \emph{extrinsic} to the electron dynamics. Hence it is difficult to formulate \emph{intrinsic} 
and universal bounds.

A recent proposal advanced in \cite{Hartnoll:2014lpa} suggested that, in analogy to the bound on 
$\eta/s$, the quantities that could be universally bounded in strange metals are the thermo-electric 
diffusion constants $D_+$ and $D_-$. More specifically in \cite{Hartnoll:2014lpa} it was argued 
that strongly correlated metals in the \emph{incoherent} regime (where momentum is quickly 
degraded) have charge and heat diffusion constants that eventually saturate the bound
\begin{equation}
\label{bound}
D_{\pm} \ge C \frac{\hbar \bar{v}^2}{k_B T},
\end{equation}
where $C$ is an undetermined constant and $\bar{v}$ is a characteristic velocity of the critical and relativistically invariant system%
\footnote{We avoid referring to $\bar v$ as a Fermi velocity strictly speaking because the latter 
does not admit in general a sharp definition in strongly correlated systems.}. 
This velocity represents an extra scale of the 
relativistic low-energy effective description which 
may be naturally related to a UV cut-off (as described for instance in \cite{Kovtun:2014nsa}).
It is relevant to note that, in the holographic context, an extra scale of the effective description
was for instance used in \cite{Blake:2014yla} in order to set two different temperature scalings 
for the resistivity and the Hall angle.

The diffusion constants $D_+$ and $D_-$ are related to the transport coefficients via 
the Einstein relations, namely:
\begin{eqnarray}\label{sum}
D_+ D_- &=& \frac{\sigma}{\chi}\, \frac{\kappa}{c_\rho}\ ,\\ \label{prod}
D_+ + D_- &=& \frac{\sigma}{\chi} + \frac{\kappa}{c_\rho} + \frac{T(\zeta \sigma - \chi s)^2}{c_\rho \chi^2 \sigma}\ ,
\end{eqnarray}
where $\sigma$, $s$ and $\kappa$ are respectively the electric, the thermo-electric and 
the thermal conductivities (see \cite{Hartnoll:2014lpa} for a derivation); 
$c_\rho$ is the specific heat at fixed charge density $\rho$, $\zeta$ is 
the thermo-electric susceptibility and $\chi$ is the electric susceptibility. 
The bounds \eqref{bound} can therefore be translated in terms of the 
thermo-electric transport coefficients of the model%
\footnote{As observed in \cite{Kovtun:2008kx}, we underline the fact that the definitions of individual 
quantities like $\sigma$ or $\eta$ are sensitive to the normalization of the associated current. 
For the purpose of formulating universal statements is then crucial to consider quantities which are 
independent of normalization choices as the ratios giving rise to the diffusivities in \eqref{sum} and \eqref{prod}.
Note that, in line with Onsager's symmetry argument about the thermo-electric conductivity matrix
(namely $s$ represents both the thermo-electric and the electro-thermal entries), the normalization 
of the heat flow is not independent of that of the electric current.}.
In particular, relying on qualitative arguments%
\footnote{Specifically on the assumption that susceptibilities are essentially 
temperature independent in the relevant regime.}, in \cite{Hartnoll:2014lpa} it was suggested 
that incoherent metals approximatively saturating the bounds \eqref{bound} have a linear-in-temperature 
resistivity controlled precisely by the equilibration time-scale $\tau_{\text{eq}} \sim \hbar/(k_B T)$.

Soon after the analysis of \cite{Hartnoll:2014lpa}, the author of \cite{Kovtun:2014nsa}, relying on 
hydrodynamical arguments, suggested that a bound on the sum of the thermoelectric diffusion constants 
is more natural than bounding the same diffusivities individually, namely
\begin{equation}
 D_+ + D_- \geq C'\frac{\hbar \bar v^2}{k_B T}\ ;
\end{equation}
still \cite{Kovtun:2014nsa} (to which we refer for details) observes that the sum $D_+ + D_-$ is always 
a real quantity and argues on the basis of a perturbative approach. 

In this paper we first study the charge and heat diffusion in the simplest holographic model featuring 
spatial diffeomorphism breaking by means of a mass term for the graviton and argue that its dynamics 
is not suitable to feature linear in $T$ resistivity and diffusion bounds. For this reason we consider 
a generalized model comprehending also a dilaton field. We first show that the phenomenology of this 
massive gravity dilaton model has interesting and robust features (among which, notably, a linear in 
$T$ resistivity) and remarkably we find that the sum of diffusion constants is bounded 
in line with \cite{Kovtun:2014nsa}.

The key point and novelty of the present analysis consists in considering the massive gravity dilaton 
model at vanishing chemical potential but with a non-trivial dilaton profile. This is essential 
to lead to linear in $T$ resistivity in the whole temperature range. Such linear in $T$ behavior
of the resistivity is robust also with respect to the momentum dissipation physics meaning that
its linearity in $T$ proves insensitive to the specific strong or weak momentum dissipation regime.

\section{The simplest holographic model: massive gravity}

Our computations rely on the phenomenologically
motivated holographic model \cite{Vegh:2013sk} which features massive gravitons in the dual 
gravitational $3+1$ dimensional bulk. The UV derivation and ultimate consistency of massive gravity
are still open questions which, to the present purposes, can be set aside as long as
the phenomenology of the $2+1$ dimensional boundary theory is itself consistent. This was proven to be the case 
both in relation to the thermodynamics \cite{Blake:2013bqa} and on the level of the thermo-electric transport 
properties \cite{Amoretti:2014zha}.

The gravitational action is
\begin{equation}
\label{action}
S = \int d^4x\,  \sqrt{-g}\, \Big[ R + \frac{6}{L^2} + \mathcal{M}(g) -\frac{1}{4 }F_{\mu \nu}F^{\mu \nu} \Big] +S_{\text{c.t.}} \ ,
\end{equation}
where  $L$ is the asymptotic $AdS_4$
radius and
\begin{equation}\label{massalphabeta}
  \mathcal{M}(g) = \alpha\; \text{tr}({\cal K}) + \beta \left[\text{tr}(\mathcal{K})^2 - \text{tr}(\mathcal{K}^2) \right]\ ,
\end{equation}
represents the graviton mass term expressed in terms of the matrix $\mathcal{K}$ whose definition is
$\mathcal{K}^\mu_{\ \nu} \equiv ( \sqrt{ \mathcal{K}^2} )^{\mu}_{\ \nu}$ with $(\mathcal{K}^2)^\mu_{\ \nu}\equiv g^{\mu \rho}f_{\rho \nu}$
and the two parameters $\alpha$ and $\beta$ having both dimension of mass$^2$.
The matrix $f$ is a non-dynamical auxiliary metric whose explicit form is
$f_{\mu \nu}= \text{diag}(0,0,1,1)$. The action \eqref{action} contains also
the Maxwell-Einstein term (with $F=dA$) and a boundary term $S_{\text{c.t.}}$ 
(see \cite{Amoretti:2014zha} for the proper definition of the latter) necessary to have a well-defined 
variational problem and a finite on-shell action. Henceforth we consider $L=1$ since this is just a 
choice of unit of measure .

Following the standard holographic dictionary, it is known that \eqref{action}
is the gravitational dual of a strongly correlated system with charged degrees of freedom exhibiting 
an extrinsic elastic mechanism of momentum dissipation such as that generated by the presence 
of quenched disorder  \cite{Vegh:2013sk,Blake:2013bqa,Davison:2013jba,Amoretti:2014zha}.
The model admits planar black brane solutions of this kind
\begin{equation}
ds^2=\frac{1}{z^2} \left[-f(z) dt^2 + dx^2 + dy^2 + \frac{1}{f(z)} dz^2\right]\ , \ \ \ \ \ 
A=a(z)\, dt\ ,
\end{equation}
where the explicit functions appearing in the ansatz are
\begin{equation}\label{ansa}
a(z) = \mu \left(1-\frac{z}{z_h} \right) \ , \ \ \ \ \ 
f(z) = 1 +z^2 \beta -\frac{z^3 \beta }{z_h} -\frac{z^3}{z_h^3} +\frac{ z \alpha }{2} -\frac{ z^3 \alpha }{2 z_h^2}
-\frac{z^3 \mu ^2}{4 z_h} + \frac{z^4 \mu ^2}{4 z_h^2}\ ,
\end{equation}
where $z_h$ is the horizon radius defined by $f(z_h)=0$ and $\mu$ is the chemical potential of the dual quantum field theory. 
According to the holographic dictionary, this black brane solution is
dual to finite temperature equilibrium states of the boundary theory whose thermodynamic
variables are encoded in the parameters and quantitative details of the bulk background, namely
\begin{equation}
\label{themodyna}
T = \frac{1}{4 \pi z_h} \left(3 + z_h \alpha + z_h^2 \beta -\frac{z_h^2 \mu ^2}{4} \right)\ , \ \ \ \ \
\mathcal{S}=\frac{4 \pi }{ z_h^2} \ , \ \ \ \ \
\rho=\frac{\mu}{ z_h}  \ ,
\end{equation}
where $T$ is the temperature, $\mathcal{S}$ is the entropy density and $\rho$ is the charge density.

The parameter $\alpha$ and $\beta$ control the strength of the momentum dissipation mechanism,
namely the momentum is dissipated faster as $|\alpha|$ and $|\beta|$ increase. 
In particular, in the coherent regime, where momentum is slowly dissipated, an extrinsic dissipation rate 
$\tau^{-1}_{\text{ext}}$ can be defined \cite{Davison:2013jba}:
\begin{equation}
\label{rate}
\tau^{-1}_{\text{ext}} = -\frac{4 (\alpha + 2 \beta  z_h)}{12 + 4 z_h \alpha + z_h^2 (4 \beta + 3 \mu^2)} \ .
\end{equation}

In order to obtain the diffusion constants from the system of equations \eqref{sum} and \eqref{prod},
we need at first the explicit expressions of the DC transport coefficients. An analytical computation (see \cite{Amoretti:2014mma} for the case $\alpha=0$)
leads to
\begin{equation}\label{transport}
\sigma = 1-\frac{z_h \mu ^2}{\alpha + 2 z_h \beta} \ , \; \; \; \; \; 
s = \frac{4 \pi  \mu }{\alpha + 2 z_h \beta } \ , \; \; \; \; \;
\kappa = -\frac{\pi  \left[z_h^2 \left(4 \beta -\mu ^2\right)+4 z_h \alpha +12\right]}{z_h^2 \left(\alpha + 2 z_h \beta -z_h   \mu ^2 \right)}\ ;
\end{equation}
secondly, we have to compute $c_{\rho}$, $\chi$ and $\zeta$ for the model at hand. 
This can be easily achieved using their thermodynamic definitions, namely
\begin{eqnarray}\label{chichi}
\chi = \left(\frac{\partial \rho}{\partial \mu}\right)_T &=& \frac{z_h^2 \left(3 \mu ^2-4 \beta \right)+12}{z_h^3 \left(\mu ^2-4
   \beta \right)+12 z_h} \ , \\ \label{zetazeta}
\zeta = \left(\frac{\partial \rho}{\partial T}\right)_\mu &=& \frac{16 \pi  \mu }{z_h^2 \left(\mu ^2-4 \beta \right)+12} \ , \\
c_\rho =  T \left(\frac{\partial \mathcal{S}}{\partial T}\right)_\mu - \frac{\zeta^2T}{\chi} &=& 
-\frac{8 \pi  \left[z_h^2 \left(4 \beta -\mu ^2\right)+4 z_h \alpha +12\right]}
{z_h^4 \left(4 \beta -3 \mu ^2\right)-12 z_h^2}\ ,
\end{eqnarray}
where, in order to perform the thermodynamic derivatives with respect to $T$ and $\mu$ one has to keep into account that $z_h$
is a function of both the temperature and the chemical potential due to the first of the relations \eqref{themodyna}.
It is interesting to note that \eqref{chichi} and \eqref{zetazeta} do not explicitly depend on $\alpha$ (they do so only through $z_h$);
formally they have the same expression as in the $\alpha = 0$ case \cite{Amoretti:2014mma}.

\section{Diffusion constants}
\label{diffmass}
\begin{figure}[ht]\label{diffu}
\begin{center}
\includegraphics[width=.45\textwidth]{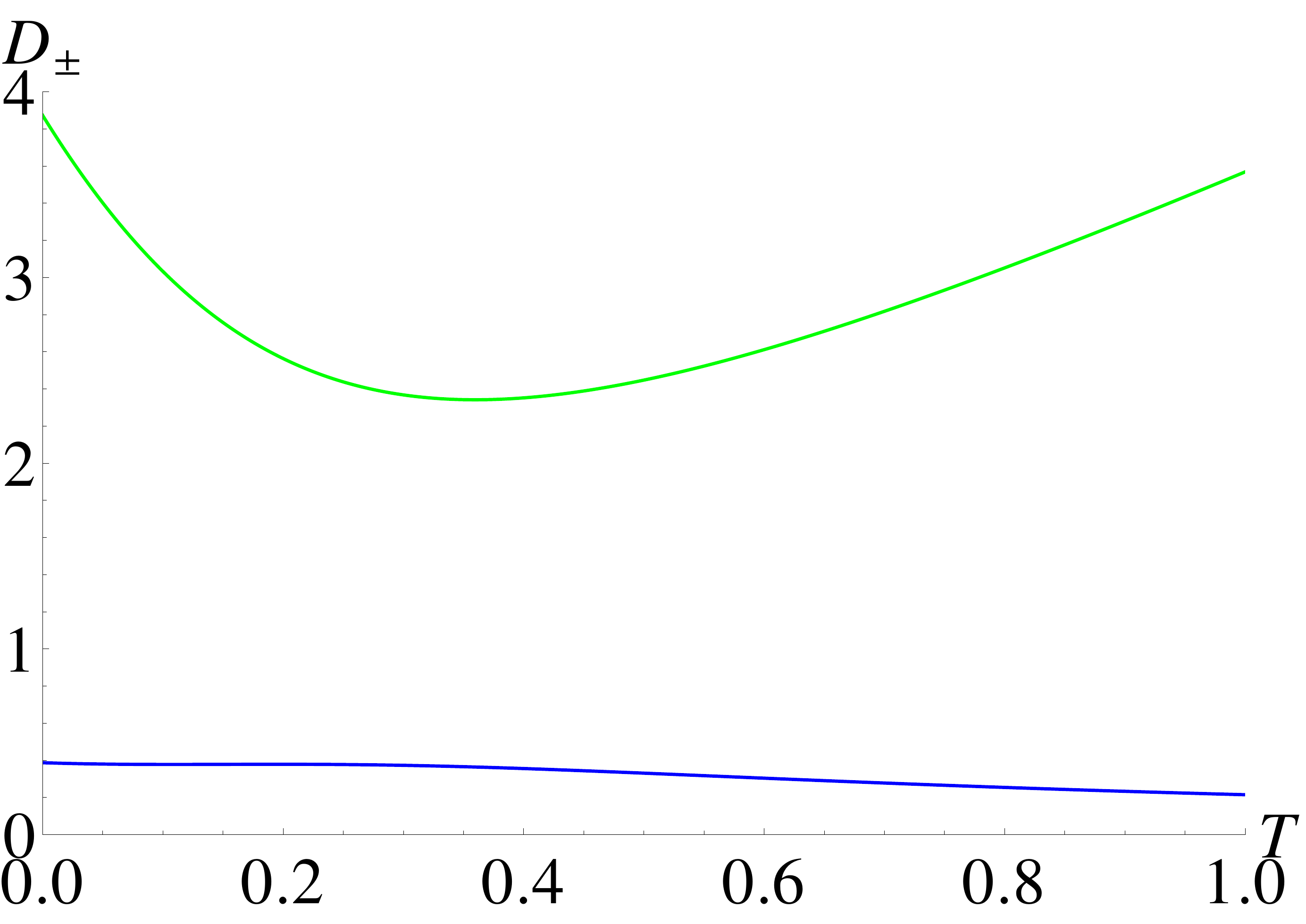}
\includegraphics[width=.45\textwidth]{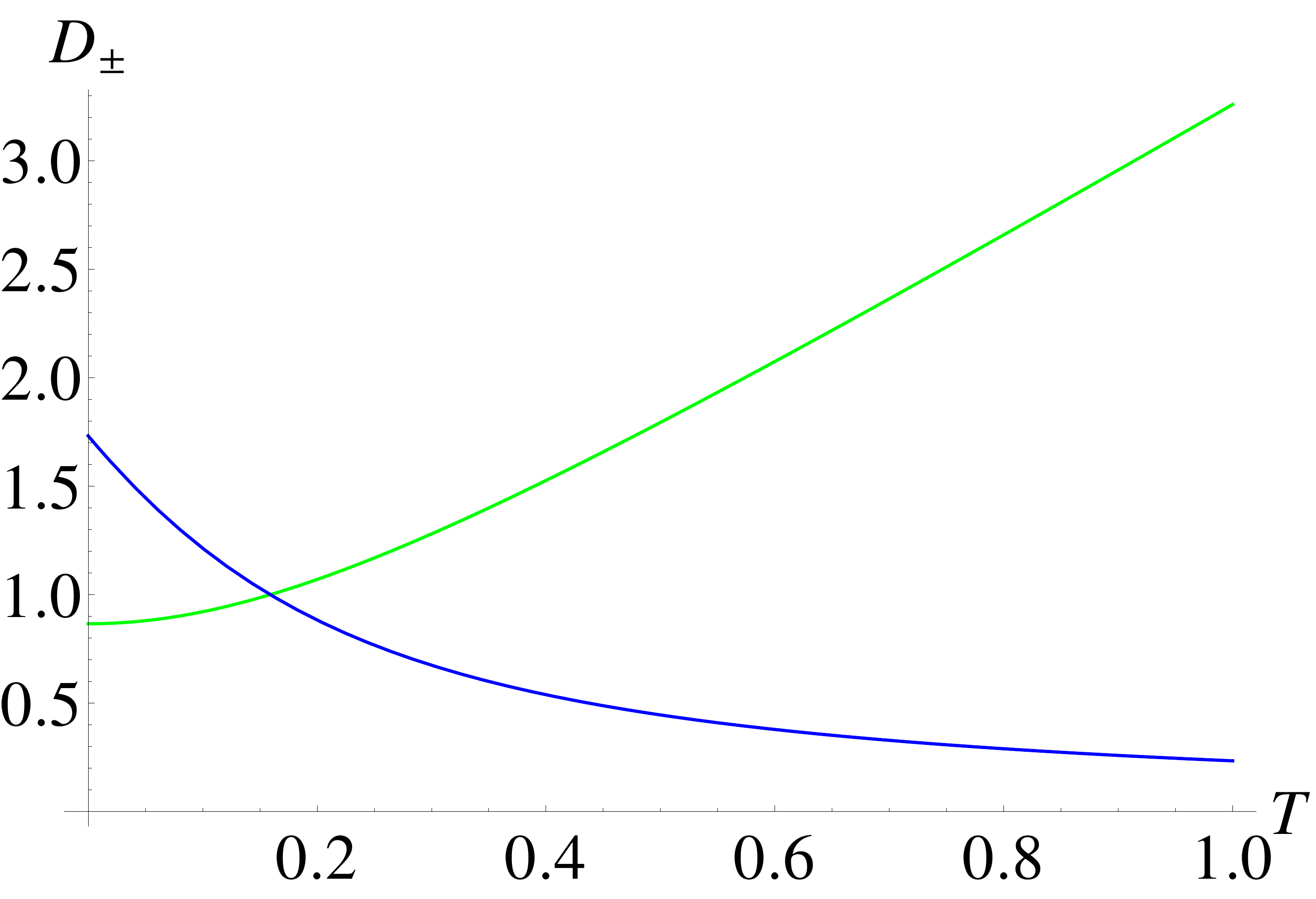}
\caption{Sample diagrams to illustrate the behavior of thermo-electric diffusion constants at 
(left) finite chemical potential, namely $\mu = 1$ (blue line for $D_-$ and green line for $D_+$), and (right) $\mu=0$ (blue line for $D_-=D_c$ and green line for $D_+=D_h$). The graviton mass parameter 
has been chosen to be $ \beta = -1$.}
\end{center}
\end{figure}

Once both the thermodynamics and the linear response of the system are under control, it is possible 
to study the diffusion constants through the Einstein relations \eqref{sum} and \eqref{prod}. 
At first we will consider the simplest case in which the parameter $\alpha$ is set to zero, 
and we will comment on the general case at the end of the section. The diffusion constants for $\alpha=0$ take the following form
\begin{equation}
D_\pm=\frac{-12+z_h^2 \left(20 \beta -19 \mu ^2\right)\pm
   \sqrt{\Delta
   }}{32 \beta  z_h}\ ,
\end{equation}
where 
\begin{equation}
\Delta=z_h^4 \left(144 \beta ^2-696 \beta  \mu ^2+361 \mu ^4\right)+z_h^2 \left(288 \beta +456 \mu ^2\right)+144\ .
\end{equation}
We refer to Figure \ref{diffu} to have a qualitative idea of the generic behaviour of the diffusion constants
with respect to the temperature. 
In the simple case of zero chemical potential $\mu = 0$ (see right panel of \ref{diffu}) the electric and thermal 
sector decouple and the two diffusivities take the following form
\begin{eqnarray}
 D_c &=& \frac{\sigma}{\chi} = - \frac{\sqrt{4 \pi^2 T^2 - 3 \beta } - 2 \pi T}{\beta} \ , \\
 D_h &=& \frac{\kappa}{c_\rho} = - \frac{\sqrt{4 \pi^2 T^2 - 3 \beta}}{\beta}\ .
\end{eqnarray}
We observe that in the high temperature regime $D_c$ has an $1/T$ behavior; this holds true 
also when $\alpha \neq 0$. As we will comment later, $\beta$ is negative.
Even though also for $\mu\neq 0$ the diffusivities are bounded from below by an $1/T$ power law 
in the high temperature region,
the coefficient of such bounding behavior can be apparently lowered at will acting on the mass
parameter $\beta$. Nevertheless, to be precise on this point, one needs to consider 
whether the range of $\beta$ is possibly limited by consistency requirements of the model in such a way to
produce a lower bound for the conductivities.

By means of simple manipulations, based on the expression for $T$ in \eqref{themodyna}, it is possible to show that
demanding a positive momentum dissipation rate $\tau^{-1}_{\text{ext}}$ \eqref{rate} (with $\alpha=0$) implies
\begin{equation}\label{re1}
 \beta  < 0\ .
\end{equation}
A further consistency requirement is provided by asking the positivity of the energy density.
More specifically, we consider the holographic renormalization of the model \eqref{action} on the black brane solutions
\eqref{ansa} and the assumption that finite counter-terms do not affect the thermodynamics
(namely that they do not depend on the thermodynamic variables) as discussed in \cite{Blake:2013bqa}.
It is then possible to refer the renormalized energy density to its value at zero $T$ (keeping fixed all the other quantities)
and ask that its value at finite $T$ be never lower than that at $T=0$, namely
\begin{equation}
\label{reen}
 {\cal E}(T,\mu;\beta) - {\cal E}(0,\mu;\beta) > 0 \ .
\end{equation}
In this $\alpha = 0 $ case, this energy requirement does not furnish any additional constraint on the parameter $\beta$, 
and the renormalized energy density appearing in \eqref{reen} remains positive for every $\beta < 0$.

All in all, the consistency requirements, although restricting the possible choices of mass parameters,
do not lead either to a diffusion bound nor to a minimal value for the diffusion constants below which the holographic model
is not trustworthy. Indeed it is always possible to achieve a strongly incoherent regime (where momentum is quickly dissipated) 
by sending $\left| \beta \right| $ to infinity without any apparent consistency problem.

In line with these observations, one has that the infinite $|\beta|$ limit leads 
to vanishing diffusion constants, namely
\begin{equation}\label{trilim}
 \lim_{\beta \rightarrow -\infty} D_{\pm}(T,\beta) = 0\ .
\end{equation}
Despite this limit corresponds to a bulk model dominated just by the mass term of the graviton 
where the dynamics of the metric vector fluctuation of interest trivializes,
such argument does not yield a quantitative ground to formulate a bound.
Even observing that an infinite $|\beta|$ limit leads necessary to a regime where the 
graviton mass exceeds the Planck mass
one could limit the value of the gravity model as an effective field theory rather than 
supporting the existence of a bound on diffusivity.

Extending to the case where $\alpha \ne 0$, the mass of the graviton is given by
\begin{equation}
m^2(z)=-2 \beta+\frac{\alpha}{z} \ ,
\end{equation}
which diverges in the UV region. We have performed the same analysis illustrated for $\alpha=0$ also for the $\alpha \ne 0$ case. 
Even though the analysis is technically more involved, there are no qualitative differences and, for every fixed values of $\alpha$,
the incoherent regime can be obtained by performing the limit $|\beta|\rightarrow \infty$ leading to unbounded diffusion constants as in \eqref{trilim}.

The outcome of the previous analyses is that the simple massive gravity model \eqref{action} does not allow to formulate consistent 
bounds on the thermo-electric diffusion constants even in the case where the mass of the graviton has non-trivial RG (i.e. radial) behavior%
\footnote{Recently a more general mass potential for the graviton has been taken into account \cite{Baggioli:2014roa}; 
even though it seems that no qualitative differences occur in this case, it would be interesting to perform here a similar analysis 
of the diffusion constants.}. 
In addition it is important to observe that the electric conductivity \eqref{transport} is, in a regime 
of large $|\beta|$, dominated by the constant term, namely the charge conjugation symmetric piece.
This highlights a limitation in reproducing a linear in $T$ resistivity in massive gravity in
a regime where momentum dissipation exceeds the other scales of the system. In other words,
the incoherent regime considered to formulate the conjecture on diffusion bounds (phenomenologically 
motivated by the consequent linear in $T$ resistivity%
\footnote{See \cite{Hartnoll:2014lpa} for further details and assumptions, such as
the insensitivity of the electric susceptibility $\chi$ to temperature.})
does not lead to the desired phenomenology in the simple massive gravity model \eqref{action} 
and, as we will see in the next sections, an analysis of the diffusion constants in a model where these requirements are fulfilled is mandatory.

\section{Adding the dilaton}

We consider a slight generalization of the massive gravity model \eqref{action}. We introduce a dilaton and consider 
the following bulk action
\begin{equation}\label{GS}
S_d=\int d^4 x \sqrt{-g} \Big[ R+6 \cosh \phi-\frac{e^{\phi}}{4}F_{\mu \nu}F^{\mu \nu}-\frac{3}{2}\partial_{\mu} \phi \partial^{\mu} \phi  
+ \mathcal{M}_\beta(g) \Big]+S_{\text{c.t.}}\ ,
\end{equation}
where the mass term we consider is given by%
\footnote{Namely the same as considered previously in \eqref{massalphabeta} fixing $\alpha = 0$.}
\begin{equation}
  \mathcal{M}_\beta(g) =  \beta \left[\text{tr}(\mathcal{K})^2 - \text{tr}(\mathcal{K}^2) \right]\ .
\end{equation}
This model is inspired by \cite{Gubser:2009qt} which was considered in $3+1$ dimensions with 
a mass term for the graviton in \cite{Davison:2013txa}. This dilaton theory is particularly appealing as it features linear in T
entropy at low temperature, characteristics which was at the core of the argument described 
in \cite{Davison:2013txa} to have linear in $T$ resistivity at low temperature ($T/\mu$). In addition, as opposed to 
\eqref{action}, we have here no residual entropy at zero temperature.

The bulk solutions of \eqref{GS} are
\begin{equation}
\begin{split}
&ds^2=\frac{g(z)}{z^2} \left(-h(z)dt^2+\frac{dz^2}{g(z)^2h(z)}+dx^2+dy^2 \right) \ ,\\
&A_t=\sqrt{\frac{3Q(Qz_h+1)}{z_h}\left(1+\frac{\beta z_h^2}{(Q z_h +1)^2}\right)}\frac{z_h-z}{z_h(Qz+1)} \ , \\
&\phi(z)=\frac{1}{3}\log g(z) \ , \qquad g(z)=(1+Qz)^\frac{3}{2} \  , \\
&h(z)=1+\frac{\beta z^2}{(Qz+1)^2} -\frac{z^3(Qz_h+1)^3}{z_h^3(Qz+1)^3}\left(1+\frac{\beta z_h^2}{(Qz_h+1)}\right) \ .
\end{split}
\end{equation}
Where the parameter $Q$ controls the dilaton profile and is related to the chemical potential.
The fact that the dilaton scalar charge is determined by the others parameters of the solution
represents a known result in the physics of dilatonic charged black holes \cite{Poletti,Gibbons:1985ac,PhysRevD.51.5720};
in other words, asymtotically $AdS$ solutions can have scalar hair however it is not independend of the other parameters 
of the black hole solution (as it happens, for instance, in \cite{Gubser:2009qt}). 
Massive gravity does not change the picture and, specifically, the argument presented in \cite{PhysRevD.51.5720} can be repeated 
in the presence of a massive graviton too.

Indeed, the thermodynamics of the model is holographically related to the bulk solution in the following way
\begin{equation}\label{ThermoGS}
\begin{split}
&T=\frac{3(1+Qz_h)^2+\beta z_h^2}{4 \pi  (1+Q z_h)^{\frac{3}{2}}z_h} \ ,  \; \; \; \mathcal{S}=\frac{4 \pi}{z_h^2}(Qz_h+1)^{\frac{3}{2}} \ , \\
&\; \; \; \; \mu=\sqrt{\frac{3Q(Qz_h+1)}{z_h}\left[1+\frac{\beta z_h^2}{(Q z_h +1)^2}\right]} \ , \\
& \qquad \qquad \qquad \rho=\frac{\mu}{z_h} (Qz_h+1) \ ,
\end{split}
\end{equation}
Relying again on analytical computation (still along the lines of a generalization of the membrane paradigm
described in \cite{Donos:2014cya}) we have analytical control upon the entire set of thermo-electric 
transport coefficients and obtain the following explicit expressions
\begin{equation}
\begin{split}
&\sigma=\frac{2 \beta  \left(Q z_h+1\right)-3 Q z_h \left[\beta +\left(\frac{1}{z_h}+Q\right)^2\right]}{2 \beta  \sqrt{Q z_h+1}} \ , \\
& s =-\frac{2 \sqrt{3} \pi  }{\beta  z_h} \sqrt{Q \left(Q z_h+1\right) \left[Q \left(Q z_h+2\right)+\beta  z_h+\frac{1}{z_h}\right]}\ , \\
&\kappa=\frac{4 \pi  \left(Q z_h+1\right) \left[3 \left(Q z_h+1\right){}^2+\beta  z_h^2\right]}{z_h^2 \left[\beta  z_h \left(Q z_h-2\right)+3 Q \left(Q
   z_h+1\right)^2\right]} \ .
\end{split}
\end{equation}
We remind the reader that, also in the present dilaton model, the positivity of the momentum dissipation rate requires $\beta < 0$
so in the preceding formul\ae~$\beta$ is always negative.

\section{Analysis at criticality}
\label{crit}

The common structure of the cuprate phase diagram (see for instance \cite{hussey}) shows that the strange metal phase is reached, in general, at high $T$.
Given the scaling properties of the holographic model with respect to a rescaling of the boundary space-time%
\footnote{See for instance \cite{Amoretti:2014zha} for the analysis of the scaling properties of the pure massive gravity model.},
it is natural to compare the temperature to the chemical potential and consider the scaling invariant ratio $T/\mu$.
Hence we mean high temperatures in the sense $T/\mu \gg 1$ and, inverting the qualitative argument just given,
it is natural to study the $\mu = 0$ case to describe the ``criticality'' condition which gives rise to the strange metal.
On top of this, when the chemical potential is vanishing, the charge and heat transport of the model decouple.

Asking for $\mu = 0$ in \eqref{ThermoGS} we face two possibilities. We can set $Q=0$ and 
consequently trivialize the dilaton profile falling back to the solution \eqref{ansa} already found in simple massive gravity. Alternatively we can set 
\begin{equation}\label{crita}
 Q z_h + 1 = |\beta|^{1/2} z_h\ ;
\end{equation}
we focus on this second case where the dilaton has non-trivial radial profile.
As we will see shortly, this is far more than a technicality. Indeed considering 
such non-trivial critical condition relates the parameter $Q$ to the other scales of the model, specifically 
the temperature (through $z_h$) and the graviton mass $\beta$. Fixing $\mu = 0$ by imposing relation \eqref{crita}
leads to a holographic model whose phenomenology differs from that of simple massive gravity. As we are going to describe 
precisely, the new features of the dilaton model lead to linear in $T$ resistivity and an overall physical behavior
in line with the that indicated in \cite{Hartnoll:2014lpa} as the basis for discussing diffusion bounds%
\footnote{We postpone the investigation of the \emph{a priori} reasons 
of such an interesting physical behavior and a possible detailed interpretation of the boundary theory dual to the
dilaton model to future work. A relevant study in this sense (although performed in massless gravity) is described in \cite{Gubser:2009qt}.}.

Studying the thermodynamics of the critical conditions associated to \eqref{crita} leads to
\begin{equation}\label{tercri}
 T = \frac{|\beta|^{1/4}}{2 \pi z_h^{1/2}} \ , \ \ \ \ \
 \rho = 0 \ , \ \ \ \ \
 {\cal S} = 8 \pi^2 |\beta|^{1/2} T\ ,
\end{equation}
which feature linear in $T$ entropy at all temperature. Moreover, regarding the 
susceptibilities, we obtain
\begin{eqnarray}\label{suscri}
 \zeta &=& 0 \ ,\\
 \chi  &=& |\beta|^{1/2} \ ,\\
 \label{ccri}
 c_\rho &=& 8 \pi^2 |\beta|^{1/2} T\ .
\end{eqnarray}
Particularly interesting to us is the electric susceptibility $\chi$ independent from 
the temperature. Continuing the analysis of the critical model resulting from the condition \eqref{crita},
we find the transport coefficients to be
\begin{eqnarray}\label{tracri}
 \sigma &=& |\beta|^{1/4} z_h^{1/2} \ \implies \sigma^{-1} = 2 \pi |\beta|^{-1/2} T \ ,\\
 s &=& 0 \ ,\\
 \kappa &=& 16 \pi^3 |\beta|^{-1/2} T^2\ .
\end{eqnarray}
We underline the fact that the resistivity is linear in $T$ for the entire range of temperature.

It is important to notice that the system does not have an independent scalar source;
as commented before, this is a general feature of the class of solutions. 
The system is therefore constrained and the scalar source is given in terms of, for instance, $\mu$ and $T$. Such 
constraint is an intrinsic propertry of the model and not an arbitrary operation or choice.

The set of equilibrium and transport quantities just found fulfills precisely the phenomenological 
framework that was considered in formulating diffusion bounds in \cite{Hartnoll:2014lpa}.
More specifically, a constant electric susceptibility was assumed as a hypothesis and a linear in 
$T$ resistivity as the consequence of an electrical diffusivity saturating the bound. 
Moreover, in \cite{Hartnoll:2014lpa} it was conjectured that the charge diffusion constant $D_c$ obeys a bound which depends only on temperature;
the arguments relying on the hypothesis that the momentum dissipation rate is fast with respect to the scale of $T$, a regime referred 
to as \emph{incoherent} regime. 

Analyzing the dilaton model we find that, in the critical condition associated to \eqref{crita}, 
the charge diffusion constant takes a very simple form independent on the parameter $\beta$, namely
\begin{equation}
\label{chargecrit}
D_c^{(\text{crit})}=\frac{\sigma}{\chi}=\frac{1}{2 \pi T} \ .
\end{equation}
It is very important to note that \eqref{chargecrit} (emerging from \eqref{tercri}- \eqref{ccri} and \eqref{tracri})
does not refer to any specific regime for the momentum dissipation rate controlled by $\beta$; in this sense, it is not
necessarily related to either an incoherent or a coherent regime. Such feature is very appealing
because allows us to be ``agnostic'' about the hierarchy of $T$ and the momentum dissipating scale $|\beta|$;
the phenomenology of the critical model at hand is therefore robust also in this sense.

Let us now focus on the heat diffusion constant $D_h$. 
The assumptions of constant susceptibility $\chi$ and linear in $T$ resistivity do not suggest a bound for $D_h$
which, still following \cite{Hartnoll:2014lpa}, can nevertheless be conjectured by analogy with $D_c$. 
For the critical dilaton model $D_h$ takes the following explicit form
\begin{equation}
\label{heatcrit}
D_h^{(\text{crit})}=\frac{\kappa}{c_{\rho}}=\frac{2 \pi T}{\left| \beta \right|} \ .
\end{equation}
In order to discuss the possibility of formulating a bound,
let us consider the incoherent regime where momentum is dissipated quickly. 
As for the simple massive gravity model considered in Section \ref{diffmass}, the incoherent regime is achieved 
in the limit $T/\left|\beta \right| \rightarrow 0$.
The heat diffusion constant \eqref{heatcrit} depends explicitly on $\beta$ while instead the charge diffusion constant \eqref{chargecrit} does not.
This crucial difference leads to the impossibility to rely on incoherence and formulate 
a lower bound which depends only on temperature for $D_h^{(\text{crit})}$.

Although one cannot formulate a bound for both the heat and charge diffusion constants separately,
as a direct consequence of the bound on $D_c^{(\text{crit})}$, the sum of the two is naturally bounded once incoherence 
is considered, namely
\begin{equation}
\label{boundss}
D_h^{(\text{crit})}+D_c^{(\text{crit})} \ge \frac{1}{2 \pi T} \ .
\end{equation}
In \cite{Kovtun:2014nsa} different arguments were proposed to motivate a bound on the sum of diffusion constants rather
than on them individually relying on a hydrodynamical analysis. We refer to \cite{Kovtun:2014nsa} for a detailed discussion on the topic but intuitively we note that the quantity $D_h+D_c$ is more natural to formulate a bound since in general the two 
diffusion constants separately could be a complex quantities while their sum is always real. 

The present analysis of the holographic dilaton model moves somehow oppositely with respect to \cite{Kovtun:2014nsa}: 
we were able to compute the two diffusion constants separately and we have found that a bound can be formulated only 
on the sum of the two once incoherence is considered.

The set of results obtained through the study of the critical massive gravity dilaton model, and specifically \eqref{boundss}, appear to be
intimately related to the detail of the model. Actually, as we have seen explicitly, simple massive gravity led to different results.
At the best of our knowledge, the dilaton model under consideration studied at $\mu = 0$ according to \eqref{crita} is the only holographic model 
were all the assumptions made in \cite{Hartnoll:2014lpa} are fulfilled. It is therefore particularly relevant that in such circumstances a bound
on the sum of the diffusion constants can be naturally formulated. In other words it is possible to consider the present analysis as a support
(at least in a specific and well defined case) of the physics conjectured in \cite{Hartnoll:2014lpa}.

\section{Discussion}
\label{disc}

In this paper we have analyzed heat and thermal diffusion constants in two different models exhibiting momentum dissipation
realized by means of a bulk massive graviton. In the simplest holographic massive gravity theory \cite{Vegh:2013sk}
we find that diffusion constants are not bounded from below and, more precisely, vanish in the incoherent limit where momentum 
is dissipated quickly. We also show that a careful analysis of the graviton mass parameter space constrained by 
physical consistency requirements (such as a positive momentum dissipation rate and a positive energy density)
does not allow to bound the diffusivities and neither to give a lower diffusion value beyond which the model becomes unreliable. 
These results are not surprising when confronted with the framework adopted by \cite{Hartnoll:2014lpa}
to propose a conjecture on diffusion bounds. Indeed, the conjecture itself is formulated in relation to
phenomenological properties like a linear in $T$ resistivity and a constant electric susceptibility 
which are not realized by the simplest massive gravity model.

To pursue the study on diffusion appears to be necessary to enrich the holographic model.
Hence we consider the addition of a dilaton field as in \cite{Davison:2013txa}. 
We show that this dilaton theory studied in specific critical conditions reproduces the desired
phenomenological behavior, $\sigma^{-1}\sim T$ and constant $\chi$, at all values of the temperature. 
This constitutes a relevant result by itself. 

We pursue the analysis of diffusion for the dilaton model to check the connection between transport features and diffusion bounds suggested 
in \cite{Hartnoll:2014lpa}. The critical dilaton model leads to a diffusion bound (of the conjectured $1/T$ form) in the charge sector while
featuring an unbounded heat diffusion constant. These picture can be connected to the study performed in \cite{Kovtun:2014nsa}
where, on general grounds, it is argued in favor on a diffusion bound on the sum of the thermo-electric diffusion constants 
rather than on them individually. 
In conclusion, the present analysis of the critical massive gravity dilaton model offers support to both the 
proposal of diffusion bounds advanced in \cite{Hartnoll:2014lpa} and \cite{Kovtun:2014nsa}.

\section{Acknowledgements}

A particular thank goes to R.~Davison
for crucial discussions on the first version of the paper.
We would also like to thank M.~Carrega, N.~ Maggiore and A.~Mezzalira. 
D.M. would also like to thank R.~Argurio, M.~Bertolini, F.~Bigazzi, P.~Creminelli, N.~Iqbal, D.~Forcella, L.~Pando Zayas, N.~Pinzani-Fokeeva, G.~Policastro, F.~Porri, D.~Redigolo
and D.~T.~Son for very nice and insightful discussions.
A.A. acknowledges support of a grant from the John Templeton foundation. 
The opinions expressed in this publication are those of the authors and do not necessarily reflect the views of the John Templeton foundation.
A. B. thanks the MIUR-FIRB2012 - Project HybridNanoDev (Grant
No. RBFR1236VV)
 and European Union FP7/2007-2013
under REA grant agreement no 630925 - COHEAT.

\end{document}